\documentclass[12pt]{article}

\usepackage{amssymb,latexsym,amsmath,amscd,color,url,framed,graphics}
\usepackage{enumerate}
\usepackage{framed}
\usepackage{graphicx}
\usepackage{eucal}
\usepackage[colorlinks]{hyperref}
\usepackage{url}
\usepackage{mathabx}
\usepackage[all]{xy}

\makeatletter

\@addtoreset{figure}{section}
\def\thefigure{\thesection.\@arabic\c@figure}
\def\fps@figure{h, t}
\@addtoreset{table}{bsection}
\def\thetable{\thesection.\@arabic\c@table}
\def\fps@table{h, t}
\@addtoreset{equation}{section}

\makeatother

\begin{document}

\newtheorem{theorem}{Theorem}[section]
\newtheorem{definition}[theorem]{Definition}
\newtheorem{lemma}[theorem]{Lemma}
\newtheorem{remark}[theorem]{Remark}
\newtheorem{proposition}[theorem]{Proposition}
\newtheorem{corollary}[theorem]{Corollary}
\newtheorem{example}[theorem]{Example}

\def\below#1#2{\mathrel{\mathop{#1}\limits_{#2}}}

%%%%%%%%%%%%%%%%%%%%%%%%%%%%%%%%%%%%%%%%%%%%%%%%%%%%%%%%%%%%%%%%%%%%%%%%%%%%%%%
%%%%%%%%

%%%%%%%%%%%%%%%%%%%%%%%%%%%%%%%%%%%%%%%%%%%%%%%%%%%%%%%%%%%%%%%%%%%%%%%%%%%%%%%
%%%%%%%%

\title{Integrable $G$-Strands on semisimple Lie groups}
\author{Fran\c{c}ois Gay-Balmaz$^{1}$, Darryl D. Holm$^{2}$,  and Tudor S. Ratiu$^{3}$}
\addtocounter{footnote}{1}
\footnotetext{CNRS / Laboratoire de 
M\'et\'eorologie Dynamique, \'Ecole Normale Sup\'erieure, 
Paris, France. Partially supported by a ÒProjet Incitatif de RechercheÓ contract from the Ecole Normale
Sup\'erieure de Paris.
\texttt{gaybalma@lmd.ens.fr}
\addtocounter{footnote}{1}}
\footnotetext{Department of Mathematics, Imperial College London. London SW7 2AZ, UK. Partially supported by the European Research Council's Advanced Grant 267382 FCCA.
\texttt{d.holm@imperial.ac.uk}
\addtocounter{footnote}{1}}
\footnotetext{Section de
Math\'ematiques and Bernoulli Center, \'Ecole Polytechnique 
F\'ed\'erale de Lausanne. CH--1015 Lausanne. Switzerland. Partially supported by Swiss NSF grant 200021-140238 and by the government grant of the Russian Federation for support of research projects implemented by leading scientists, Lomonosov Moscow State University under  agreement No. 11.G34.31.0054. 
\texttt{tudor.ratiu@epfl.ch}
\addtocounter{footnote}{1} }

\date{PACS: 02.20.Sv, 02.30.Ik, 02.30.Jr}
\maketitle

\makeatother
%\begin{center} DRAFT \end{center}
\maketitle

%|||-------------------text width----------------------|||

%\noindent \textbf{AMS Classification:}

%\noindent \textbf{Keywords:}

\begin{abstract}
The present paper derives systems of partial differential equations that admit a quadratic zero curvature representation for an arbitrary real semisimple Lie algebra. It also determines the general form of  Hamilton's principles and Hamiltonians for these systems and analyzes the linear stability of their equilibrium solutions in the examples of $\mathfrak{so}(3)$ and $\mathfrak{sl}(2,\mathbb{R})$.
\end{abstract}

\tableofcontents

\section{Introduction}

The principal chiral model is a map $c:\mathbb{R}^{1,1}\to G$ from the Minkowski space-time $\mathbb{R}^{1,1}$ into a Lie group $G$. For its relation to harmonic maps into Lie groups $h:\mathbb{R}^{2}\to G$, see \cite{DuFoNo1992,Ma1994,Uh1989}. Moreover, the principal chiral model is also an integrable system of partial differential equations (PDEs) {whose} doubly infinite sequence of conservation laws was found in \cite{Di1983a} and whose soliton solutions can be generated using the \textit{dressing method}, related to the classical Riemann-Hilbert problem in complex analysis. Both the conservation laws and the dressing method have been deduced from a \textit{zero curvature representation} (ZCR)  of the PDE system for the chiral model  
\cite{DuFoNo1992,DuKrNo1985,FaTa1987,FoTr1988,NoMaPiZa1984,ZaMi1978,ZaMi1980}. 
One of the key ideas behind the derivation of the conservation laws \cite{Di1983a} is the use of Hamilton's principle. 

Suppose $G$ is a semisimple Lie group. Then the PDEs for the principal chiral model may be derived from Hamilton's principle $\delta S = 0$, with $S=\iint \ell(u,v)ds\,dt$, where the symmetry-reduced Lagrangian $\ell:\mathfrak{g}\times \mathfrak{g}  \to \mathbb{R}$ is defined using the ad-invariant Cartan-Killing form on the Lie algebra $\mathfrak{g}$ of the semisimple Lie group $G$.  {For semisimple Lie groups, the resulting PDEs may be} expressed only in terms of Lie bracket operations, which in turn lead to the ZCR for {these systems}.  The ZCR for the principal chiral model involves matrix differential operators with \textit{rational} combinations of the spectral parameter that leads to the inverse scattering transform method and the dressing method for these PDEs. 

It is also possible to obtain Hamiltonian structures and Lax equations  for the principal chiral model generated by using matrix differential operators with polynomial dependence on a parameter \cite{Di1983b}.
For example, a recent paper \cite{Ge2013} derives new types of integrable 4-wave equations via a Lie algebra approach for $\mathfrak{so}(5,\mathbb{C})$. These equations were derived directly from a zero curvature representation (ZCR) that is quadratic in a spectral parameters and were shown to admit solutions that may be constructed via the Riemann-Hilbert problem. Polynomial dependence of ZCRs on a parameter was also discussed in \cite{Ge2012}. 

The present paper derives PDE systems that admit a quadratic ZCR for an arbitrary real semisimple Lie algebra and determines the general form of their Hamilton's principles and Hamiltonians. Several examples are given, including the cases of $\mathfrak{so}(3)$, $\mathfrak{sl}(2,\mathbb{R})$, $\mathfrak{so}(4)$, and $\mathfrak{g}_2$. In all cases, the Hamiltonians are quadratic in the dependent variables, but sign-indefinite, so they admit instabilities. {The effects of these instabilities on the full nonlinear solution behavior remains to be understood, perhaps through numerical simulations.}

Our approach is to derive the equations from Hamilton's principle for a Lagrangian defined on a Lie algebra $\mathfrak{g}$ that is invariant under transformations by the corresponding Lie group $G$. This approach produces $G$-Strand equations \cite{HoIv2013, HoIvPe2012}. A $G$-Strand is a map $g:\,\mathbb{R}\times\mathbb{R}\to G$ for a Lie group $G$ that follows from Hamilton's principle for a certain class of $G$-invariant Lagrangians by using the Euler-Poincar\'e (EP) theory explained in \cite{Ho2011}. The $SO(3)$-strand may be regarded physically as a continuous strand of rigid frames, as for a spin chain \cite{Ho2011,HoIvPe2012}. The simplest example of the $SO(3)$-strand is the case in which the moment of inertia is proportional to the identity. This case recovers the principal chiral model, which as we know admits a ZCR that is not of the quadratic type we study in this note \cite{Ma1994}. Here, we use the EP theory to derive the $G$-Strand equations for any semisimple Lie algebra, then we derive the quadratic ZCR for such a system.  

\paragraph{Plan of the paper.}
In section \ref{EP-GStrands-sec} we discuss the $G$-Strand equations for an arbitrary semisimple Lie algebra. In section \ref{chiral-sec} we discuss the chiral model for $SO(3)$. In section \ref{ZCRs-sec} we construct ZCRs that are quadratic in a spectral parameter for two classes of equations. These classes follow from either a normal real form in section \ref{case1-sec}, or a compact real form in section \ref{case2-sec}. These constructions impose additional relations among the dependent variables directly in the equations, in order for them to admit a quadratic ZCR. This type of substitution of functional relations would not in general produce a Hamiltonian system. However, the expressions for the conserved energy in the cases derived from Hamilton's principle and EP theory provided clues for how to derive $G$-Strand PDEs that are manifestly Hamiltonian and admit the same ZCR. In particular, these equations possess an affine Lie--Poisson bracket defined on the dual of the corresponding Lie algebra $\mathfrak{g}$ of the Lie group $G$.

\section{$G$-Strand equations}
\label{EP-GStrands-sec}
\paragraph{Lagrangian formulation.} Let $G$ be a real Lie group with
Lie algebra $\mathfrak{g}$. The $G$-Strand equations are
\begin{equation}
\label{GS_eqn}
\partial _t \frac{\delta \ell}{\delta \xi }- \operatorname{ad}^*_ \xi \frac{\delta \ell}{\delta \xi }+ \partial _s \frac{\delta \ell}{\delta \gamma }- \operatorname{ad}^*_ \gamma \frac{\delta \ell}{\delta \gamma }=0, \quad  \partial _t \gamma - \partial _s \xi +[ \xi , \gamma ]=0,
\end{equation}
where $\ell: \mathfrak{g}\times \mathfrak{g} \rightarrow \mathbb{R}$ and $ \xi , \eta : \mathbb{R}  \times \mathbb{R}  \rightarrow \mathfrak{g}  $.
This system of PDEs follows from Hamilton's principle
\[
\delta \iint \ell(\xi(t,s), \gamma(t,s)) ds \,dt=0
\] 
for the Lagrangian $\ell(\xi,\gamma)$ with $\xi=g^{-1}g_t$ and 
$\gamma=g^{-1}g_s$, for the map $g:\mathbb{R}\times\mathbb{R}\to G$. The first of these $G$-Strand equations depends on the choice of Lagrangian $\ell$, while the second one does not. That is, the first one is dynamic and the second one is kinematic.

When $\mathfrak{g}$ admits a non-degenerate bi-invariant bilinear symmetric form, one can use it to identify the dual space $ \mathfrak{g}  ^\ast $ with $ \mathfrak{g}  $. With this identification, we have $ \operatorname{ad}^*_ \xi= - \operatorname{ad}_ \xi $ and the $G$-Strand equations \eqref{GS_eqn}  become in this case
\begin{equation}
\label{GS_kappa_eqn}
\partial _t \frac{\delta \ell}{\delta \xi }+\operatorname{ad}_ \xi \frac{\delta \ell}{\delta \xi }+ \partial _s \frac{\delta \ell}{\delta \gamma }+ \operatorname{ad}_ \gamma \frac{\delta \ell}{\delta \gamma }=0, \quad  \partial _t \gamma - \partial _s \xi +[ \xi , \gamma ]=0.
\end{equation}

We will use the notations
\[
\mu := \frac{\delta \ell}{\delta \xi }, \quad 
\pi := \frac{\delta \ell}{\delta \gamma },
\]
so that the $G$-Strand equations \eqref{GS_kappa_eqn} read
\begin{equation}\label{G_strand} 
\partial _t \mu +[ \xi , \mu ]+ \partial _s \pi +[ \gamma , \pi ]=0, \quad \partial _t \gamma - \partial _s \xi +[ \xi , \gamma ]=0.
\end{equation}

\paragraph{Hamiltonian formulation.} Let $\mathfrak{g}^\ast$ be the 
dual vector space to $\mathfrak{g}$ and $\left\langle\,, \right\rangle:
\mathfrak{g}^\ast\times \mathfrak{g} \rightarrow \mathbb{R}$ the natural
non-degenerate duality pairing. Define the Hamiltonian function
$h : \mathcal{F} ( \mathbb{R}  ,\mathfrak{g})^\ast \times \mathcal{F} ( \mathbb{R} , \mathfrak{g}) \rightarrow\mathbb{R}$ by
$h(\mu, \gamma) : = \int \left( \left\langle \mu, \xi \right\rangle - \ell( \xi, 
\gamma) \right) ds$, where $\mu \in \mathfrak{g}^\ast$ and $\xi \in \mathfrak{g}$
is such that $\frac{\delta \ell}{\delta \xi} = \mu$. At this point,
we assume that $\ell$ is non-degenerate in $\xi$ which guarantees that
$\frac{\delta \ell}{\delta \xi} = \mu$ has a unique solution $\xi(\mu)$.

We have the relations
\[
\frac{\delta h}{\delta \mu }= \xi 
\quad\text{and}\quad 
\frac{\delta h}{\delta \gamma }=- \frac{\delta \ell}{\delta \gamma},
\]
so equations \eqref{GS_kappa_eqn} read 
\begin{equation}
\label{GS_Ham}
\partial _t \mu +\operatorname{ad}_ {\frac{\delta h}{\delta \mu  }} \mu - \partial _s \frac{\delta h}{\delta \gamma }- \operatorname{ad}_ \gamma \frac{\delta h}{\delta \gamma }=0
\quad\text{and}\quad 
\quad  \partial _t \gamma - \partial _s \frac{\delta h}{\delta \mu}  +\left[ \frac{\delta h}{\delta \mu } , \gamma \right] =0
\,.
\end{equation}
They belong to the class of the affine Lie--Poisson equations \cite{GBRa2009}.
Note that equations \eqref{GS_Ham} do not require the existence of a
Lagrangian $\ell$. As we shall see later on, this is an important
remark in the formulation of the general $G$-Strand equations.

From the physical viewpoint, the $G$-Strand equations \eqref{GS_Ham} are exactly Hamilton's equations for a static ideal complex fluid whose broken symmetry is the semisimple Lie group $G$ \cite{Ho2002,GBRa2009,
EGBHPR2010}.
Thus, an additional motivation for studying the $G$-Strands is the hope that integrable systems may be found that apply for the description of nonlinear waves on the order parameter spaces of complex fluids in the absence of dissipation.

\section{Chiral model} \label{chiral-sec}

One of the best studied $G$-Strand equations is given by the chiral model which we now briefly recall.

For a complex Lie algebra $\mathfrak{g}^\mathbb{C}$, let 
$u, v: \mathbb{R}^2 \rightarrow \mathfrak{g}^\mathbb{C}$ be smooth 
functions satisfying the \textit{principal chiral field equations} 
(see, e.g., \cite{Ma1994})
\begin{equation}
\label{chiral_equ}
\partial_yu - \frac{1}{2}[u,v] = 0, \qquad 
\partial_xv + \frac{1}{2}[u,v] = 0.
\end{equation}
This system is equivalent to
\begin{equation}
\label{chiral_equ_second}
\partial_yu - \partial_xv - [u,v] = 0,\qquad 
\partial_y u + \partial_xv = 0.
\end{equation}
The first equation implies that locally there exist a smooth map
$g: U\subset \mathbb{R}^2 \rightarrow G^\mathbb{C}$, $U$ open 
and simply connected in $\mathbb{R}^2$, $G^\mathbb{C}$ the connected
simply connected complex Lie group with Lie algebra $\mathfrak{g}^\mathbb{C}$,  such that 
$u =  g^{-1}(\partial_x g)$ and $v = g^{-1}(\partial_y g)$ (see
\cite[Chapter V, Theorem 2.4]{Sternberg1983}) and hence the second
equation is equivalent to
\[
\partial_y\left(g^{-1}(\partial_x g)\right) + 
\partial_x\left(g^{-1}(\partial_y g)\right) = 0.
\]
If $\mathfrak{g}^\mathbb{C}$ admits a non-degenerate bi-invariant symmetric
bilinear form $\kappa$, these equations admit a $G$-Strand formulation 
\eqref{GS_kappa_eqn} with $\ell(u,v): = \kappa(u,v)$.

The spacetime coordinates for the chiral model are given by 
$t= \frac{1}{2}(x-y)$, $s= \frac{1}{2}(x+y)$. Then, $\xi: = 
g^{-1}(\partial_tg)$ and $\gamma: = g^{-1}(\partial_xg)$ are 
expressed in terms of $u$ and $v$ as $\xi = u-v $ and 
$\gamma = u +v$. Equations \eqref{chiral_equ_second} become
\begin{equation}
\label{chiral_t_s_equ}
\partial_t \gamma - \partial_s \xi + [\xi, \gamma]= 0,
\qquad \partial_s\gamma - \partial_t \xi = 0.
\end{equation}
They are $G$-Strand equations for $\ell(\xi, \gamma) : = 
\frac{1}{4}\left(\kappa(\gamma, \gamma) - \kappa( \xi, \xi) \right)$.
As it is well known, equations \eqref{chiral_equ_second} are equivalent to the zero curvature condition for the connection one-form
$\omega(\lambda) = \frac{u}{1-\lambda}dx + \frac{v}{1+\lambda} dy \in 
\Omega^1(\mathbb{R}^2; \mathfrak{g})$, for all $\lambda \in 
\overline{\mathbb{C}}\setminus \{\pm 1\}$ (see \cite{Ma1994}). Of course,
the same can be said about the equations \eqref{chiral_t_s_equ}.

\paragraph{Example: Chiral model for $G=SE(3)$.} The real Lie algebra 
$\mathfrak{g}  = \mathfrak{se}(3)$ of the special Euclidean group has
the following non-degenerate symmetric bi-invariant form: 
\[
\left\langle ( \boldsymbol{\Omega}_1  , \boldsymbol{\Gamma}_1  ), ( \boldsymbol{\Omega} _2 , \boldsymbol{\Gamma} _2 ) \right\rangle := \boldsymbol{\Omega} _1 \cdot \boldsymbol{\Gamma}_2 + \boldsymbol{\Gamma} _1 \cdot \boldsymbol{\Omega}_2, \qquad \boldsymbol{\Omega}_1  , \boldsymbol{\Gamma}_1, \boldsymbol{\Omega}_2  , \boldsymbol{\Gamma}_2
\in \mathbb{R}^3.
\]
So, relative to this pairing, we get
\[
\operatorname{ad}^*_{( \boldsymbol{\Omega} , \boldsymbol{\Gamma} )}( \boldsymbol{\Pi} , \boldsymbol{\Sigma} ) = - \operatorname{ad}_{( \boldsymbol{\Omega} , \boldsymbol{\Gamma} )}( \boldsymbol{\Pi} , \boldsymbol{\Sigma} )=- (\boldsymbol{\Omega} \times \boldsymbol{\Pi} ,\boldsymbol{\Omega} \times \boldsymbol{\Sigma} -  \boldsymbol{\Pi} \times \boldsymbol{\Gamma} ).
\]
The chiral equations \eqref{chiral_equ_second} become
\[
\left\{
\begin{aligned}
&\partial_y\mathbf{u}_1 - \partial_x \mathbf{v}_1 - \mathbf{u}_1 \times 
\mathbf{v}_1 = {\bf 0}\\
&\partial_y\mathbf{u}_2 - \partial_x \mathbf{v}_2 - 
\mathbf{u}_1 \times \mathbf{v}_2 + \mathbf{v}_1 \times \mathbf{u}_2 = {\bf 0}\\
&\partial_y\mathbf{u}_i + \partial_x \mathbf{v}_i = {\bf 0}, \quad i=1,2.
\end{aligned}
\right.
\]
They are $G$-Strand equations for the Lagrangian $\ell(\mathbf{u}_1, \mathbf{u}_2, \mathbf{v}_1, \mathbf{v}_2) : = \mathbf{u}_1\cdot \mathbf{v}_2 + 
\mathbf{u}_2\cdot \mathbf{v}_1$.

In spacetime coordinates $(t,s)$, these equations become
\[
\left\{
\begin{aligned}
&\partial_t\boldsymbol{\gamma}_1 - \partial_s \boldsymbol{\xi}_1 - \boldsymbol{\xi}_1 \times 
\boldsymbol{\gamma}_1 = {\bf 0}\\
&\partial_t\boldsymbol{\gamma}_2 - \partial_s \boldsymbol{\xi}_2 + \boldsymbol{\xi}_1 \times \boldsymbol{\gamma}_2 - 
\boldsymbol{\gamma}_1 \times \boldsymbol{\xi}_2 = {\bf 0} \\
&\partial_s\boldsymbol{\gamma}_i - \partial_t \boldsymbol{\xi}_i = 
{\bf 0}, \quad i=1,2.
\end{aligned}
\right.
\]
These are $G$-Strand equations for $\ell(\boldsymbol{\xi}_1, \boldsymbol{\xi}_2, \boldsymbol{\gamma}_1,\boldsymbol{\gamma}_2) :=
\frac{1}{2}\left(\boldsymbol{\gamma}_1\cdot \boldsymbol{\gamma}_2 - 
\boldsymbol{\xi}_1 \cdot \boldsymbol{\xi}_2\right)$. \quad 
$\blacklozenge$

\section{Quadratic zero curvature representations} 
\label{ZCRs-sec}

Let $\mathfrak{g}$ be a real Lie algebra with a bi-invariant 
bilinear symmetric  non-degenerate form. We consider the Lie
algebra valued functions on an open connected set 
$U \subset \mathbb{R}^2$ given by
\[
L(t,s):= \lambda ^2 a + \lambda \mu (t,s)- \gamma (t,s)\in \mathfrak{g}  
, \quad 
M(t,s):= \lambda ^2 b - \lambda \pi(t,s) - \xi(t,s) \in \mathfrak{g},
\]
where $a, b \in \mathfrak{g}  $ are constant Lie algebra elements. The associated zero curvature representation (ZCR) is defined as \cite{Ge2012,Ge2013}
\[
\partial _t L- \partial _s M+[L , M]=0
\,.
\]
Imposing the ZCR yields the following system of relations
\begin{equation}
\label{g_strand_conditions}
\left\{
\begin{aligned} 
\lambda ^0 :&\;\; \partial _t \gamma - \partial _s \xi +[ \xi , \gamma ]=0\\
\lambda ^1 :&\;\; \partial _t \mu + \partial _s \pi  + [ \xi , \mu ]+[ \gamma , \pi ]=0\\
\lambda ^2 :&\;\; [a, \xi ]+ [ \mu  , \pi ]+ [ \gamma ,b]=0\\
\lambda ^3 :&\;\; [a, \pi ]+[b, \mu ]=0\\
\lambda ^4 :&\;\; [a, b]=0.
\end{aligned}
\right.
\end{equation} 

Let us suppose that the Lie algebra $ \mathfrak{g}$ is the 
normal real form of a complex semisimple Lie algebra 
$\mathfrak{g}^\mathbb{C}$; hence $\mathfrak{g}$ is split. 
The Killing form 
$\kappa: \mathfrak{g}^\mathbb{C} \times \mathfrak{g}^\mathbb{C}
\to\mathbb{C}$ defined by 
\[
\kappa(x, y) := {\rm tr}({\rm ad}_\xi\,{\rm ad}_\eta), \qquad 
\xi, \eta\in \mathfrak{g}^\mathbb{C},
\]
is bilinear symmetric bi-invariant and non-degenerate. We shall use 
it to identify the dual space $\mathfrak{g}^\ast$ of the normal
real form with $\mathfrak{g}$ itself. From bi-invariance, using this identification, 
we have $ \operatorname{ad}^*_ \xi =- \operatorname{ad}_\xi$, 
for all $\xi\in \mathfrak{g}$. 

We recall some of the definitions and theorems from the theory
of semisimple Lie algebras (see, e.g., \cite{Humphreys1978}). 
Choose a Cartan subalgebra $\mathfrak{c}^\mathbb{C}$ of 
$\mathfrak{g}^\mathbb{C}$ and let 
\[
\mathfrak{g}^\mathbb{C} = \mathfrak{c}^\mathbb{C} \oplus
\coprod_{\alpha \in  \Delta_+} \left(\mathfrak{g}^\mathbb{C}_\alpha 
\oplus \mathfrak{g}^\mathbb{C}_{-\alpha} \right)
\]
be the associated root space decomposition, where $\Delta \subset 
\left(\mathfrak{c}^\mathbb{C} \right)^\ast$ is the set of roots.
Choose a base $\Pi:=\{\alpha_1, \ldots, \alpha_r\}$ of $\Delta$ 
relative to which one defines
the positive and negative roots $\Delta_+$ and $\Delta_-$. Recall
that $\dim_\mathbb{C} \mathfrak{g}^\mathbb{C}_{\pm \alpha} = 1$.
By non-degeneracy, each $\alpha \in \Delta$ defines a unique element
$t_\alpha\in \mathfrak{c}^\mathbb{C}$ such that $\left\langle \alpha, 
\xi\right\rangle = \kappa(t_\alpha, \xi)$, for all 
$\xi\in \mathfrak{g}^\mathbb{C}$, which in turn induces
a positive definite inner product, also denoted by $\kappa$, on
$\operatorname{span}_\mathbb{R} \Delta$, i.e., $\kappa(\alpha, \beta)
=\kappa(t_\alpha, t_ \beta)$ for all $\alpha, \beta\in \Delta$.
Given $\alpha\in \Delta$, define $h_\alpha: = 
\frac{2 t_\alpha}{\kappa(\alpha, \alpha)}$ and let 
$h_i: = h_{\alpha_i}$.

Many computations become easier if one chooses a Chevalley basis
of $\mathfrak{g}^\mathbb{C}$. This is a vector space basis
$\left\{h _i , e_\alpha \mid i=1,...,r,\; h_i \in 
\mathfrak{c}^\mathbb{C},\, \alpha \in 
\Delta,\; e_\alpha \in \mathfrak{g}_\alpha\right\}$  of $\mathfrak{g}^\mathbb{C}$ satisfying (\cite{Humphreys1978})
\begin{itemize}
\item  $[e_\alpha , e_{-\alpha }]= h_\alpha$ for all 
$\alpha\in \Delta$ ;
\item if $ \alpha , \beta , \alpha + \beta \in \Delta $, then the
constants $N_{\alpha, \beta} \in \mathbb{Z}$ defined by
$[e_\alpha , e_\beta ]=N_{\alpha, \beta }e_{ \alpha + \beta }$, 
satisfy $N_{ \alpha , \beta }=-N_{ -\alpha ,- \beta }$. 
\end{itemize}
In what follows we impose the convention $N_{ \alpha , \beta }=0$ if 
$\alpha + \beta \notin \Delta$. Such Chevalley bases always exist.

Using the appropriate Chevalley basis, the equations for the $G$-Strand can
be formulated explicitly for any semisimple Lie algebra.

\subsection{Case 1: normal real form} 
\label{case1-sec}

Given a complex semisimple Lie algebra $\mathfrak{g}^\mathbb{C}$,
its normal real form is defined in terms of a chosen Chevalley 
basis by $\mathfrak{g}: = 
\operatorname{span}_\mathbb{R}\left\{h_i, e_\alpha\mid 
i=1, \ldots, r,\, \alpha \in \Delta\right\}$, i.e., we have
\[
\mathfrak{g} =\mathfrak{c} \oplus \coprod_{\alpha \in \Delta_+} 
\left(\mathfrak{g}_\alpha \oplus \mathfrak{g}_{-\alpha} \right),
\]
where $\mathfrak{c}: = \operatorname{span}_\mathbb{R}\left\{h_i 
\mid i=1, \ldots, r\right\}$ and $\mathfrak{g}_\alpha =
\mathbb{R}e _\alpha$.

\paragraph{Lagrangian formulation.} Define the Lagrangian
\begin{equation}
\label{lagrangian_normal}
\ell_{a,c, r}( \xi , \gamma )= - \frac{1}{2} \kappa \left( \varphi _{c,a}( \xi - r \gamma ), \xi - r \gamma \right)- \kappa (c,\gamma), \quad r \in \mathbb{R}  , \quad a,c \in \mathfrak{c},
\end{equation}
where the \textit{sectional operator} (see \cite[Chapter 2]{FoTr1988}),
\[
\varphi _{c,a}= \operatorname{ad}_c ^{-1} \operatorname{ad}_a  : \mathfrak{g}   \rightarrow \mathfrak{n} _- \oplus \mathfrak{n} _+,
\]
is given by
\[
\varphi _{ c, a}( \xi )=\sum_{\alpha \in \Delta_+}  \frac{\left\langle \alpha, a \right\rangle}
{\left\langle \alpha , c\right\rangle} 
(\xi _\alpha e_{\alpha}+ \xi _{- \alpha }e_{-\alpha})
\]
with $c$ a regular semisimple element (i.e., $\left\langle\alpha, c \right\rangle\neq 0$ for all $\alpha \in \Delta$). The last three
 conditions in \eqref{g_strand_conditions}, that is,
\begin{equation}
[a,b]=0, 
\quad \left[ a, \frac{\delta \ell}{\delta \gamma }\right] + 
\left[  b, \frac{\delta \ell}{\delta \xi }\right] =0,
\quad\hbox{and}\quad 
[ a, \xi ]+ [\gamma , b]+ \left[ \frac{\delta \ell}{\delta \xi }, \frac{\delta \ell}{\delta \gamma } \right] =0
\label{ZCR-conds}
\end{equation}
are easily seen to hold.
In order to satisfy the first equation, we choose $ b = ra$. Using the expressions
\[
\mu := \frac{\delta \ell}{\delta \xi }= - \varphi _{c,a}( \xi - r \gamma )
\quad\hbox{and}\quad
\pi :=  \frac{\delta \ell}{\delta \gamma }= r \varphi _{c,a}( \xi - r \gamma )-c
\]
for the momenta, shows that the second and third equations in \eqref{ZCR-conds} are both verified.

The energy associated to the Lagrangian \eqref{lagrangian_normal} is
\[
e(\xi, \gamma) : = \kappa\left( \frac{\delta \ell}{\delta \xi}, 
\xi\right) - \ell(\xi, \gamma) = 
- \frac{1}{2} \kappa \left( \varphi _{c,a}( \xi - r \gamma ), 
\xi - r \gamma \right) - \kappa \left(
\varphi_{c,a}(\xi - r \gamma ), r \gamma \right)
+ \kappa (c,\gamma).
\]
Note that the Lagrangian \eqref{lagrangian_normal} is degenerate. This
is easily seen by noting that the Legendre transformation $\xi \mapsto
\mu: = \frac{\delta \ell}{\delta \xi}$ is not surjective since the 
Cartan subalgebra is not contained in its range. 

\paragraph{Equations.} We can rewrite the $G$-Strand equations 
\eqref{G_strand} in terms of the new variables 
$ \zeta := \xi - r \gamma$  and $\gamma$ as follows
\begin{equation}\label{new_form}
\left(\partial_t - r \partial_s\right)
\varphi_{c,a}(\zeta) +
[\zeta , \varphi _{c,a}(\zeta )]+[\gamma , c]=0, \qquad
\left(\partial_t - r \partial_s\right) \gamma - \partial_s\zeta +[ \zeta, \gamma]=0.
\end{equation}
Note that each term of the first equation is in 
$\mathfrak{n}_- \oplus \mathfrak{n}_+$,  that the second
equation in this system is independent of the chosen Lagrangian,
and that it has non-trivial projections on the Cartan subalgebra 
$\mathfrak{c}$ and on $\mathfrak{n}_- \oplus \mathfrak{n}_+$. Change
the spacetime coordinates $(t, s)$ to $(\tau, \sigma): = (t, rt+s)$
and keep the same notations for the functions $\zeta$ and $\gamma$.
Equations \eqref{new_form} become
\begin{equation}
\label{hyp_system_g_strand}
\begin{bmatrix}
\partial_\tau \varphi_{c,a}& 0\\
- \partial_\sigma& \partial_\tau
\end{bmatrix}
\begin{bmatrix}
\zeta\\ \gamma
\end{bmatrix} + 
\begin{bmatrix}
[\zeta , \varphi _{c,a}(\zeta )]+[\gamma , c]\\
[\zeta, \gamma]
\end{bmatrix}=
\begin{bmatrix}
0\\0
\end{bmatrix},
\end{equation}
or
\[
\partial_\tau
\begin{bmatrix}
\varphi_{c,a}(\zeta)\\ \gamma
\end{bmatrix} + \partial_\sigma
\begin{bmatrix}
0\\- \zeta
\end{bmatrix} + 
\begin{bmatrix}
[\zeta , \varphi _{c,a}(\zeta )]+[\gamma , c]\\
[\zeta, \gamma]
\end{bmatrix}=
\begin{bmatrix}
0\\0
\end{bmatrix}.
\]
The unknowns in this system are $(\zeta, \gamma)$. However, the
system above has only $2\dim\mathfrak{g} - \operatorname{rank}\mathfrak{g}$ equations. The missing $\operatorname{rank}\mathfrak{g}$
equations are due to the degeneracy of the Lagrangian \eqref{lagrangian_normal}.

Below, we shall define a Hamiltonian with the property that the
associated $G$-Strand equations verify the zero curvature equations
and the resulting system has $2 \dim \mathfrak{g}$ equations.

\paragraph{Hamiltonian approach.} Given $ r \in \mathbb{R}  $ and $a,c \in \mathfrak{c} $ with $a$ regular (i.e., $\left\langle\alpha, a \right\rangle \neq0 $ for all $\alpha\in \Delta$), we consider the Hamiltonian
\[
h( \mu , \gamma )= \int\left(- \frac{1}{2} \kappa ( \varphi _{a,c}( \mu ), \mu )+ \kappa (r \mu +c, \gamma )\right)ds,
\]
whose variational derivatives are 
\[
\xi =\frac{\delta h}{\delta \mu }= -\varphi _{a,c}( \mu )+r \gamma \quad\text{and}\quad \frac{\delta h}{\delta \gamma }= r \mu +c = - \pi
\,.
\]
The corresponding affine Lie--Poisson equations \eqref{GS_Ham} are
\begin{equation*}
\left\{
\begin{aligned}
&\partial _t \mu - r\partial _s \mu +[-\varphi _{a,c}( \mu )+r \gamma , \mu ]- [ \gamma , r \mu +c]=0, \\
& \partial _t \gamma - \partial _s (-\varphi _{a,c}( \mu )+r \gamma)+[-\varphi _{a,c}( \mu )+r \gamma, \gamma ]=0,
\end{aligned}
\right.
\end{equation*}
and these simplify to
\begin{equation}
\label{g_strand_hamiltonian_version}
\left\{
\begin{aligned}
&(\partial _t - r\partial _s )\mu -[\varphi _{a,c}( \mu ) , \mu ]- [ \gamma , c]=0, \\
& (\partial _t -r \partial _s )\gamma + \partial _s \varphi _{a,c}( \mu )-[ \varphi _{a,c}( \mu ), \gamma ]=0.
\end{aligned}
\right.
\end{equation}
Upon setting $b=ra$, $ \xi = -\,\varphi _{a,c}( \mu )+r \gamma $ and $ \pi = - \,r \mu -c$, we see that the ZCR conditions \eqref{g_strand_conditions} are satisfied.  Note that this system
has $2 \dim \mathfrak{g}$ equations.

\paragraph{Example: $ \mathfrak{g}  = \mathfrak{sl}(2, \mathbb{R})$.} The standard Cartan subalgebra for $\mathfrak{sl}(2, \mathbb{R})$ is 
\[
\mathfrak{c} =\left\{ a= \left.\begin{bmatrix}A& \;\;0\\ 0& -A \end{bmatrix}\,\right|\,  \in \mathbb{R}  \right\} 
\]
and the root vectors are
\[
e_\alpha = \begin{bmatrix}0& 1\\ 0&0 \end{bmatrix}, \quad e_{-\alpha} = \begin{bmatrix}0& 0\\ 1&0 \end{bmatrix},
\]
where the root $ \alpha\in \mathfrak{c} ^\ast $ is given by
\[
\left\langle  \alpha , \begin{bmatrix}A& \;\;0\\ 0& -A \end{bmatrix}\right\rangle =2A.
\]
Given $a,c \in \mathfrak{c} $ with $C\neq 0$, and writing an arbitrary Lie algebra element as
\[
\xi =\begin{bmatrix}\xi_{\mathfrak{c}} & \xi _\alpha \\ 
\xi _{- \alpha }&-\xi_{\mathfrak{c}} \end{bmatrix}
\]
we have
\[
\varphi _{c,a}\begin{bmatrix}
\xi_{\mathfrak{c}} & \xi _\alpha \\ 
\xi _{- \alpha }&-\xi_{\mathfrak{c}}  \end{bmatrix} = \frac{A}{C}\begin{bmatrix}0& \xi _\alpha \\ \xi _{- \alpha }&0 \end{bmatrix},\quad \text{for all} \quad  \xi \in \mathfrak{sl}(2, \mathbb{R}  ).
\]
The Killing form is $ \kappa ( \xi , \eta )= 4 \operatorname{Tr}( \xi \eta )$. This yields the Lagrangian
\[
\ell_{c, a, r}(\xi , \gamma )= - \frac{4A}{C}( \xi _\alpha - r \gamma _\alpha )( \xi _{-\alpha} - r \gamma _{-\alpha })  
-8 C \gamma_{\mathfrak{c}}.
\]
The first $SL(2, \mathbb{R})$-strand equation in \eqref{new_form} becomes
\[
\frac{A}{C}\begin{bmatrix}0& \partial _t \zeta  _\alpha  \\ \partial _t \zeta  _{ - \alpha } &0 \end{bmatrix}- \frac{rA}{C} \begin{bmatrix}0& \partial _s \zeta  _\alpha  \\ \partial _s \zeta  _{ - \alpha } &0\end{bmatrix}+ 2 \zeta_{\mathfrak{c}} \begin{bmatrix}0&  \zeta  _\alpha  \\ -\zeta  _{ - \alpha } &0 \end{bmatrix}+ 2 C 
\begin{bmatrix}0&  -\gamma   _\alpha  \\ \gamma   _{ - \alpha } &0 \end{bmatrix},
\]
i.e.,
\[
\left\{ 
\begin{array}{l}
\vspace{0.2cm}\displaystyle\frac{A}{C}( \partial _t -r \partial _s) \zeta _\alpha + 2\zeta_{\mathfrak{c}}  \zeta _\alpha -2C \gamma _\alpha=0 \\
\displaystyle \frac{A}{C}( \partial _t -r \partial _s) \zeta _{-\alpha }+ 2\zeta_{\mathfrak{c}} \zeta _{-\alpha }+2C\gamma _{-\alpha}=0. 
\end{array}
\right .
\]
The second $SL(2, \mathbb{R})$-strand equation in \eqref{new_form} becomes
\[
\left\{ 
\begin{array}{l}
\vspace{0.2cm}(\partial_t -r \partial_s) 
\gamma_\alpha - \partial_s\zeta_\alpha + 2 \left(\zeta_{\mathfrak{c}}\gamma_\alpha - 
\zeta_\alpha \gamma_{\mathfrak{c}}\right) = 0\\
\vspace{0.2cm} (\partial_t -r \partial_s)
\gamma_{-\alpha } - \partial_s \zeta_{-\alpha} + 2 \left(
\zeta_{-\alpha} \gamma_{\mathfrak{c}} - \zeta_{\mathfrak{c}}
\gamma_{-\alpha}\right)=0\\
(\partial_t -r \partial_s)\gamma_{\mathfrak{c}} - 
\partial_s\zeta_{\mathfrak{c}} + 
\zeta_\alpha \gamma_{-\alpha} - \gamma_\alpha\zeta_{-\alpha} = 0.
\end{array} 
\right.
\]
As noted in the general theory, the Lagrangian approach is degenerate and supplies only five equations for six unknowns: there is no equation for $\zeta_{\mathfrak{c}}$.

We now pass to the Hamiltonian approach. The equations \eqref{g_strand_hamiltonian_version}
become in this case
\begin{equation*}
\left\{
\begin{aligned}
&(\partial_t - r \partial_s) \mu_{\mathfrak{c}} = 0\\
&(\partial_t - r \partial_s) \mu_\alpha + \frac{2C}{A}\mu_{\mathfrak{c}}
\mu _\alpha + 2C\gamma_\alpha = 0\\ 
&(\partial_t - r \partial_s) \mu_{-\alpha} - \frac{2C}{A}\mu_{\mathfrak{c}}
\mu _{-\alpha} -2C\gamma_{-\alpha} = 0
\end{aligned}
\right.
\end{equation*}
and 
\begin{equation*}
\qquad\quad  \left\{
\begin{aligned}
&(\partial_t - r \partial_s)\gamma_{\mathfrak{c}} - \frac{C}{A}\left(
\mu_\alpha\gamma_{-\alpha} - \gamma_\alpha\mu_{- \alpha}\right) = 0\\
&(\partial_t - r \partial_s)\gamma_\alpha +\frac{C}{A} \partial_s\mu_\alpha -\frac{2C}{A}\mu_\alpha\gamma_{\mathfrak{c}} = 0\\
&(\partial_t - r \partial_s)\gamma_{-\alpha} +\frac{C}{A} \partial_s
\mu_{-\alpha} -\frac{2C}{A}\mu_{-\alpha}\gamma_{\mathfrak{c}} = 0.
\qquad \blacklozenge
\end{aligned}
\right. 
\end{equation*}

\color{black}
\paragraph{Example: $\mathfrak{g}=
\mathfrak{g}_2(\mathbb{R})$.} To illustrate the versatility of our approach, we compute the $G$-Strand equations for the real normal form $\mathfrak{g}_2(\mathbb{R})$ of the $14$ dimensional exceptional
complex Lie algebra $\mathfrak{g}_2(\mathbb{C})$.
Using the standard root
space decomposition in the construction of 
$\mathfrak{g}_2(\mathbb{C})$ from its Dynkin diagram 
(see, e.g., \cite[\S 19.3]{Humphreys1978}), the fact that 
$\mathfrak{g}_2(\mathbb{C})$ is a Lie subalgebra of 
$\mathfrak{so}(7, \mathbb{C})$, and the isomorphism
realizing the elements of $\mathfrak{so}(7, \mathbb{C})$ as
skew-symmetric matrices
(see \cite[Chapter VIII]{Cahn1984}), one finds that the general 
$\mathfrak{g}_2(\mathbb{C})$ matrix has the following 
expression
\begin{equation}
\label{g_2}
\begin{bmatrix}
A^{\rm skew} - \frac{1}{2}\widehat{\mathbf{u}}&
{\rm i} \mathbf{v}&
-{\rm i} \left(A^{\rm sym} + \frac{1}{2}\widehat{\mathbf{v}}\right)\\
- {\rm i} \mathbf{v}^\mathsf{T}&0& - \mathbf{u}^\mathsf{T}\\
{\rm i} \left(A^{\rm sym} - \frac{1}{2}\widehat{\mathbf{v}}\right)
&\mathbf{u} &
A^{\rm skew} + \frac{1}{2}\widehat{\mathbf{u}}
\end{bmatrix} \in \mathfrak{g}_2(\mathbb{C}) \subset 
\mathfrak{so}(7, \mathbb{C}),
\end{equation}
where $\mathbf{u}, \mathbf{v} \in \mathbb{C}^3$, 
$A \in \operatorname{sl}(3, \mathbb{C})$, 
$A^{\rm skew}: = (A - A^\mathsf{T})/2$, and
$A^{\rm sym}: = (A + A^\mathsf{T})/2$. The Cartan 
subalgebra consists of matrices of this form with
$A = \operatorname{diag}(a_1, a_2 - a_1, - a_2)$
and $\mathbf{u}= \mathbf{v}= {\bf 0}$. Note that
if we have two elements of the form \eqref{g_2} 
corresponding to $(A_i, \mathbf{u}_i, \mathbf{v}_i)$, 
$i=1,2$, then the Lie bracket is of the form \eqref{g_2},
where
\begin{equation}
\label{g_2_bracket}
\left\{
\begin{aligned}
A&=[A_1, A_2] + \frac{3}{4}\left(\mathbf{u}_2 
\mathbf{v}_1^\mathsf{T} +\mathbf{v}_1\mathbf{u}_2^\mathsf{T}
- \mathbf{v}_2 \mathbf{u}_1^\mathsf{T} - 
\mathbf{u}_1\mathbf{v}_2^\mathsf{T}\right) +
\frac{3}{4} \widehat{\mathbf{u}_1 \times \mathbf{u}_2} -
\frac{3}{4} \widehat{\mathbf{v}_1 \times \mathbf{v}_2} \\
& \quad
+\frac{1}{2}(\mathbf{v}_2 \cdot \mathbf{u}_1
-\mathbf{v}_1 \cdot \mathbf{u}_2 ) \mathbf{I}_3\\
\mathbf{u} &= \mathbf{u}_1 \times \mathbf{u}_2 + 
\mathbf{v}_1 \times \mathbf{v}_2 - 
A_1^{\rm sym}\mathbf{v}_2 + A_1^{\rm skew}\mathbf{u}_2 + 
A_2^{\rm sym}\mathbf{v}_1 - A_2^{\rm skew}\mathbf{u}_1\\
\mathbf{v} &= \mathbf{u}_2 \times \mathbf{v}_1 + 
\mathbf{v}_2 \times \mathbf{u}_1 + 
A_1^{\rm skew}\mathbf{v}_2 - A_1^{\rm sym}\mathbf{u}_2 - 
A_2^{\rm skew}\mathbf{v}_1 + A_2^{\rm sym}\mathbf{u}_1
\end{aligned}
\right.
\end{equation}
and the trace of their product equals
\begin{equation}
\label{trace_g_2}
-3 \mathbf{u}_1\cdot\mathbf{u}_2 + 
3 \mathbf{v}_1\cdot\mathbf{v}_2 + 
2\operatorname{Tr}(A_1A_2). 
\end{equation}

The associated real normal form is obtained from this 
expression by setting all matrices and vectors real 
(but the factors of ${\rm i}=\sqrt{-1}$ in \eqref{g_2} remain). If $a,c$ are in
the Cartan subalgebra and $c$ is regular semisimple, the
sectional operator has the following effect on an element
of the form \eqref{g_2}
\begin{align*}
A\left[a_{ij} \right]&\longmapsto A_{c,a}: =
\begin{bmatrix}
\vspace{2mm}
0& \displaystyle{\frac{2a_1-a_2}{2c_1-c_2}a_{12}}&
\displaystyle{\frac{a_1+a_2}{c_1+c_2} a_{13}}\\
\vspace{2mm}
\displaystyle{\frac{2a_1-a_2}{2c_1-c_2}a_{21}}&0&
\displaystyle{\frac{2a_2-a_1}{2c_2-c_1}a_{23}}\\
\displaystyle{\frac{a_1+a_2}{c_1+c_2} a_{31}}&
\displaystyle{\frac{2a_2-a_1}{2c_2-c_1}a_{32}}&0 
\end{bmatrix}\\
\mathbf{u}&\longmapsto \mathbf{u}_{c,a}: =
\begin{bmatrix}
\vspace{2mm}
\displaystyle{\frac{a_1}{c_1} u_1}\\
\vspace{2mm}
\displaystyle{\frac{a_2-a_1}{c_2-c_1}u_2}\\
\displaystyle{\frac{a_2}{c_2}u_3}
\end{bmatrix}, \quad\!\! \!
\mathbf{v}\longmapsto \mathbf{v}_{c,a}: = 
\begin{bmatrix}
\vspace{2mm}
\displaystyle{\frac{a_1}{c_1} v_1}\\
\vspace{2mm}
\displaystyle{\frac{a_2-a_1}{c_2-c_1}v_2}\\
\displaystyle{\frac{a_2}{c_2}v_3}
\end{bmatrix}.
\end{align*}
The denominators do not vanish precisely because $c$ is
a regular semisimple element. Since the Killing form $\kappa$
for $\mathfrak{g}_2(\mathbb{R})$ is a constant multiple of the trace
of the product of matrices of the form \eqref{g_2}, we
shall take this coefficient to be one.

The Lagrangian \eqref{lagrangian_normal} becomes in this case
\begin{align*}
\ell_{c,a,r}(\xi, \gamma)&= \frac{3}{2} \left(
\frac{a_1}{c_1}u_1^2 + \frac{a_2-a_1}{c_2-c_1}u_2^2 +
\frac{a_2}{c_2}u_3^2 \right)
-\frac{3}{2} \left(
\frac{a_1}{c_1}v_1^2 + \frac{a_2-a_1}{c_2-c_1}v_2^2 +
\frac{a_2}{c_2}v_3^2 \right) \\
& \quad -\frac{2a_1-a_2}{2c_1-c_2}a_{21}a_{12} -
\frac{a_1+a_2}{c_1+c_2} a_{31}a_{13} - 
\frac{2a_2-a_1}{2c_2-c_1}a_{32}a_{23}\\
& \quad -2\left(c_1A^\gamma_{11}+(c_2-c_1)A^\gamma_{22} - 
c_2A^\gamma_{33}\right)
\end{align*}
where $\xi-r \gamma$ has the form \eqref{g_2} and the
matrices $A$ in \eqref{g_2} for $\gamma$ and $c$ are 
denoted by $\left[A^\gamma_{ij}\right]$ and 
$\left[A^c_{ij}\right] \in \mathfrak{sl}(3, \mathbb{R})$,
respectively. Using \eqref{g_2_bracket} and noting that
the last summand of the first equation vanishes in this
case, the first 
equation in \eqref{new_form} becomes 
\begin{align*}
&\left(\partial_t - r \partial_s \right)A_{c,a} + 
\left[A,A_{c,a} \right] + \frac{3}{4} \left(
\mathbf{u}_{c,a}\mathbf{v}^\mathsf{T} + 
\mathbf{v}\mathbf{u}_{c,a}^\mathsf{T} -
\mathbf{v}_{c,a}\mathbf{u}^\mathsf{T} -
\mathbf{u}\mathbf{v}_{c,a}^\mathsf{T} \right) \\
& \qquad + 
\frac{3}{4}\left(\widehat{\mathbf{u} \times\mathbf{u}_{c,a}}
-\widehat{\mathbf{v} \times\mathbf{v}_{c,a}} \right) + 
\left[A ^\gamma, A^c \right] = 0\\
& \left(\partial_t - r \partial_s \right)\mathbf{u}_{c,a} + 
\mathbf{u} \times\mathbf{u}_{c,a} + 
\mathbf{v} \times\mathbf{v}_{c,a} -
A^{\rm sym} \mathbf{v}_{c,a} + A^{\rm skew} \mathbf{u}_{c,a}
+ A_{c,a}^{\rm sym} \mathbf{v} -
A_{c,a}^{\rm skew} \mathbf{u} =0\\
& \left(\partial_t - r \partial_s \right)\mathbf{v}_{c,a} + 
\mathbf{u}_{c,a} \times\mathbf{v} + 
\mathbf{v}_{c,a} \times\mathbf{u} +
A^{\rm skew} \mathbf{v}_{c,a} - A^{\rm sym} \mathbf{u}_{c,a}
- A_{c,a}^{\rm skew} \mathbf{v} +
A_{c,a}^{\rm sym} \mathbf{u} =0.
\end{align*}
The explicit form of the second equation in \eqref{new_form} can
be easily obtained by decomposing each of the terms 
according to \eqref{g_2} and using the bracket formula \eqref{g_2_bracket}.

As noted in the general theory, the Lagrangian approach to the $G$-Strand equations misses two equations. The Hamiltonian approach corrects this problem. The  equations can be obtained, as above, by decomposing $\mu$ and $\gamma$ according to \eqref{g_2} and using the expressions of the Lie bracket in \eqref{g_2_bracket}. \quad $\blacklozenge$

\subsection{Case 2: compact real form} \label{case2-sec}

Fix a Chevalley basis of $\mathfrak{g}^\mathbb{C}$. Define the \textit{compact real form} of 
$\mathfrak{g}^\mathbb{C}$ by
\begin{equation}
\label{compact_real_form}
\begin{aligned} 
\mathfrak{l}: &=\left\{\left.{\rm i}\sum_{j=1}^r a_j h_j + 
\sum_{\alpha \in \Delta_+} x_\alpha(e_\alpha- e_{-\alpha}) +
{\rm i}\sum_{\alpha \in \Delta_+} y_\alpha(e_\alpha+ 
e_{-\alpha}) \;\right|\; a_j, x_\alpha, y_\alpha\in\mathbb{R}
\right\}\\
&=:{\rm i}  \mathfrak{c} \oplus \mathfrak{u} \oplus \mathfrak{v}.
\end{aligned} 
\end{equation} 
The Lie algebra $\mathfrak{l}$ is real and compact (i.e.,
the Killing form $\kappa$ is negative definite on 
$\mathfrak{l}$), ${\rm i}\mathfrak{c}$ is
its Cartan subalgebra, $\mathfrak{l} \otimes_{\mathbb{R}} 
\mathbb{C} = \mathfrak{g}^\mathbb{C}$,
and $\mathfrak{l}$ is the fixed point set of the anticomplex
involution $\sigma: \mathfrak{g}^\mathbb{C}\rightarrow 
\mathfrak{g}^\mathbb{C}$ given in the Chevalley basis by
$\sigma(h_j) = -h_j$ and $\sigma(e_\alpha) = - e_{-\alpha}$
for all $j=1, \ldots, r$, $\alpha \in \Delta$. For example,
if $\mathfrak{g}^\mathbb{C} = \mathfrak{sl}(r+1, \mathbb{C})$,
then $\sigma(\xi) = - (\bar{\xi})^\mathsf{T}$ for all
$\xi\in \mathfrak{sl}(r+1, \mathbb{C})$, and $\mathfrak{l} = 
\mathfrak{su}(r+1)$.

Given $ a \in \mathfrak{c} $ and writing $ \xi = \sum_{ \alpha \in \Delta _+}( x _\alpha u _\alpha + y _\alpha v _\alpha) $, we have the formulas
\[
\operatorname{ad}_{{\rm i}a} \xi = \sum_{ \alpha \in \Delta _+} \left\langle \alpha , a \right\rangle (- y _\alpha u _\alpha + x _\alpha v _\alpha )\quad\text{and}\quad \operatorname{ad}_{{\rm i}a}^{-1}  \xi=   \sum_{ \alpha \in \Delta _+} \frac{1}{\left\langle \alpha , a \right\rangle} ( y _\alpha u _\alpha - x _\alpha v _\alpha ),
\]
where, in the second equality, $ a \in \mathfrak{c} $ is regular semisimple.

Given $ a, b \in  \mathfrak{c} $, we have
\[
\operatorname{ad}_{{\rm i}a}^{-1} \operatorname{ad}_{{\rm i}b} \left( {\rm i} h+ \sum_{ \alpha \in \Delta _+}( x _\alpha u _\alpha + y _\alpha v _\alpha) \right) =  \sum_{ \alpha \in \Delta _+}\frac{\left\langle \alpha , b \right\rangle }{\left\langle \alpha , a \right\rangle } ( x _\alpha u _\alpha + y _\alpha v _\alpha) 
\]

Let us define the Lagrangian
\[
\ell_{a,c, r}( \xi , \gamma )= - \frac{1}{2} \kappa \left( \varphi _{c,a}( \xi - r \gamma ), \xi - r \gamma \right)- \kappa ({\rm i} c,\gamma), \quad r \in \mathbb{R}  , \quad a,c \in \mathfrak{c},
\]
with the sectional operator (see \cite[Chapter 2]{FoTr1988}),
\[
\varphi _{c,a}= \operatorname{ad}_{{\rm i} c} ^{-1} \operatorname{ad}_{ {\rm i} a } : \mathfrak{u} \oplus \mathfrak{v} \rightarrow \mathfrak{u}  \oplus \mathfrak{v},
\]
where $c$ is regular semisimple. Using the expressions
\[
\mu = \frac{\delta \ell}{\delta \xi }= - \varphi _{c,a}( \xi - r \gamma ), \quad\pi =  \frac{\delta \ell}{\delta \gamma }= r \varphi _{c,a}( \xi - r \gamma )-{\rm i} c\,,
\]
it is easy to check that $\ell$ verifies the conditions
\[
[a,b]=0, \quad \left[ a, \frac{\delta \ell}{\delta \gamma }\right] + \left[  b, \frac{\delta \ell}{\delta \xi }\right] =0, \quad [ a, \xi ]+ [\gamma , b]+ \left[ \frac{\delta \ell}{\delta \xi }, \frac{\delta \ell}{\delta \gamma } \right] =0,
\]
with $b=ra$.

\paragraph{Equations.} As in the case of the real normal form, 
we can rewrite the equations 
\eqref{G_strand} in terms of the new variables 
$ \zeta := \xi - r \gamma$  and $\gamma$ as follows
\begin{equation}\label{new_form2}
\left(\partial_t - r \partial_s\right)
\varphi_{c,a}(\zeta) +
[  \zeta , \varphi _{c,a}(  \zeta )]+[ \gamma , {\rm i}c]=0, \qquad
\left(\partial_t - r \partial_s\right) \gamma - \partial_s\zeta +[ \zeta, \gamma]=0.
\end{equation}
Note that each term of the first equation is in 
$\mathfrak{u} \oplus \mathfrak{v}$. Note that the second
equation in this system is independent of the chosen
Lagrangian but contains, as in the normal real form, components in the Cartan algebra.

\paragraph{Hamiltonian approach.} Given $ r \in \mathbb{R}  $ and 
${\rm i}a, {\rm i}c \in \mathfrak{c}$ with $a$ regular, we consider the Hamiltonian
\begin{equation}
\label{compact_hamiltonian}
h(\mu , \gamma )= \int \left(-\, \frac{1}{2} \kappa ( \varphi _{a,c}( \mu ), \mu )+ \kappa (r \mu+ {\rm i}c, \gamma )\right)ds.
\end{equation}
We have
\[
\xi =\frac{\delta h}{\delta \mu }= -\,\varphi _{a,c}( \mu )+r \gamma \quad\text{and}\quad \frac{\delta h}{\delta \gamma }= r \mu + {\rm i}c.
\]
The affine Lie--Poisson equations are
\begin{equation*}
\left\{
\begin{aligned}
&\partial _t \mu - r\partial _s \mu +[-\varphi _{a,c}( \mu )+r \gamma , \mu ]- [ \gamma , r \mu +{\rm i}c]=0, \\
& \partial _t \gamma - \partial _s (-\varphi _{a,c}( \mu )+r \gamma)+[-\varphi _{a,c}( \mu )+r \gamma, \gamma ]=0,
\end{aligned}
\right.
\end{equation*}
and simplify to
\begin{equation}
\label{g_strand_hamiltonian_version_compact}
\left\{
\begin{aligned}
&(\partial _t - r\partial _s )\mu -[\varphi _{a,c}( \mu ) , \mu ]- [ \gamma , {\rm i}c]=0, \\
& (\partial _t -r \partial _s )\gamma + \partial _s \varphi _{a,c}( \mu )-[ \varphi _{a,c}( \mu ), \gamma ]=0.
\end{aligned}
\right.
\end{equation}

With $b=ra$, $ \xi = -\varphi _{a,c}( \mu )+r \gamma $ and $ \pi = - r \mu -{\rm i}c$, we see that \eqref{g_strand_conditions} are satisfied.

\paragraph{Example: $\mathfrak{g}  = \mathfrak{so}(3)$.}  The
linear map $\mathbb{R}^3\ni \mathbf{u} \longmapsto\widehat{\mathbf{u}} \in \mathfrak{so}(3)$, where $\widehat{\mathbf{u}} \mathbf{v}: = 
\mathbf{u}\times \mathbf{v}$, for any $\mathbf{v}\in\mathbb{R}^3$,  
is a Lie algebra isomorphism between
$(\mathbb{R}^3, \times )$ and $(\mathfrak{so}(3), [\,,])$, i.e.,
$\left[\widehat{\mathbf{u}}, \widehat{\mathbf{v}} \right] = 
\widehat{ \mathbf{u} \times \mathbf{v}}$. Note that 
$\operatorname{tr}\left(\widehat{\mathbf{u}} \widehat{\mathbf{v}}\right)
= -2 \mathbf{u}\cdot \mathbf{v}$. Below we shall identify these two 
Lie algebras.

The Lie 
algebra $\mathfrak{so}(3) \cong \mathbb{R}^3$ is the compact real form of the complex 
simple Lie algebra $\mathfrak{so}(3, \mathbb{C})\simeq \mathbb{C}^3$. 
Recall that the complex dimension of any Cartan subalgebra is one, 
i.e., $r=1$ in the general theory. From the definition \eqref{compact_real_form} of the compact real form of 
$\mathfrak{so}(3, \mathbb{C})$, once the real Cartan subalgebra 
$\mathfrak{c}$ is chosen, there exists a unique, up to real scaling, 
non-zero  $\mathbf{A} \in \mathbb{R}^3$, such that 
${\rm i}\mathfrak{c}
= \operatorname{span}_\mathbb{R}\mathbf{A}$. Therefore, any 
$\mathbf{a}, {\bf c} \in \mathfrak{c}$ are of the form 
$\mathbf{a}= {\rm i}a \mathbf{A}$ and ${\bf c} = {\rm i}c \mathbf{A}$, $a, c \in \mathbb{R}$. The associated sectional 
operator is hence
\[
\varphi _{\mathbf{a} , \mathbf{c}}(\boldsymbol{\zeta}) =
\operatorname{ad}_{{\rm i}\mathbf{a}}^{-1}
\operatorname{ad}_{{\rm i}{\bf c}} (\boldsymbol{\zeta}) 
= \frac{c}{a}\left(\boldsymbol{\zeta}  - \left(\boldsymbol{\zeta}\cdot
\frac{\mathbf{A}}{\|\mathbf{\mathbf{A}}\|}\right)
\frac{\mathbf{A}}{\|\mathbf{\mathbf{A}}\|} \right)  
= \frac{c}{a}\boldsymbol{\zeta}_{\perp} ,
\]
where $\boldsymbol{\zeta}_{\perp}$ is the projection of 
$\boldsymbol{\zeta}$ on the subspace $\mathbf{A}^\perp$. 
The Hamiltonian \eqref{compact_hamiltonian} becomes
\begin{align}
\begin{split}
h(\boldsymbol{\mu}, \boldsymbol{\gamma})&= \int\left(\frac{c}{a}\boldsymbol{\mu}_{\perp} \cdot 
\boldsymbol{\mu} -2(r\boldsymbol{\mu} +c\mathbf{A})\cdot 
\boldsymbol{\gamma}\right)ds
\\
&= \int \left(\frac{c}{a}\|\boldsymbol{\mu}_\perp\|^2 -
2r\boldsymbol{\mu} \cdot \boldsymbol{\gamma} - 
2c \mathbf{A}\cdot \boldsymbol{\gamma} \right)ds 
\end{split}
\label{SO3Ham}
\end{align}
and hence, relative to the Killing form (so minus twice the dot product)
\[
\frac{\delta h}{\delta \boldsymbol{\mu}}=
- \frac{c}{a}\boldsymbol{\mu}_{\perp}+r \boldsymbol{\gamma}, \qquad\frac{\delta h}{\delta \boldsymbol{\gamma} }= r\boldsymbol{\mu} + 
c\mathbf{A}.
\]
The affine Lie--Poisson equations  \eqref{g_strand_hamiltonian_version_compact} are
\begin{equation}
\label{two_so3_zcr_equn}
(\partial _t -r \partial _s )\boldsymbol{\mu} - 
\frac{c}{a}\boldsymbol{\mu}_{\perp} \times\boldsymbol{\mu} - 
\boldsymbol{\gamma} \times c \mathbf{A}=0 \quad\text{and}\quad 
(\partial _t -r \partial _s ) \boldsymbol{\gamma} + 
\frac{c}{a}\partial_s \boldsymbol{\mu}_{\perp} - 
\frac{c}{a} \boldsymbol{\mu}_{\perp} \times\boldsymbol{\gamma} =0.
\end{equation}

\noindent\textsf{Physical interpretation of ZCR $SO(3)$-strands.} For 
$\mathfrak{so}(3)$, the quadratic ZCR $G$-Strand equations describe nonlinear 
unidirectional vector waves on a type of spin chain that is analogous to a strand of heavy tops strung along a filament with arc length $s$.
The quadratic Hamiltonian \eqref{SO3Ham} is not positive definite, which means that, in general, its equilibria will not be stable. In particular, it affords no control over either $\|\boldsymbol{\gamma}\|$ or $\|\boldsymbol{\mu}\|$.

An alternative form of the affine Lie--Poisson equations is 
\begin{equation}
\frac{\partial}{\partial t}
    \begin{bmatrix}
    \boldsymbol{\mu}
    \\
    \boldsymbol{\gamma}
    \end{bmatrix}
=
\begin{bmatrix}
 \boldsymbol{\mu}\times
   &
   \partial_s + \boldsymbol{\gamma}\times
   \\
    \partial_s + \boldsymbol{\gamma}\times
   & 0
    \end{bmatrix}
    \begin{bmatrix}
   \delta h/\delta \boldsymbol{\mu} \\
   \delta h/\delta \boldsymbol{\gamma} 
    \end{bmatrix}
    ,
    \label{LP-Ham-struct-SO3}
\end{equation}
which makes their analogy with the heavy top very clear. Namely, we are dealing with heavy tops of angular frequency $\boldsymbol{\xi}
=\frac{c}{a} \boldsymbol{\mu}_{\perp}$ at each location $s$, spread out along a filament and interacting with their neighbors through their gradients in $s$. Equations \eqref{two_so3_zcr_equn} give two unidirectional vector wave equations.

One may verify directly that these equations imply the following characteristic-derivative relations
\begin{equation}
\label{cons-eqns}
\left\{
\begin{aligned}
(\partial _t -r \partial _s ) 
\left(\frac{c}{a} \|\boldsymbol{\mu}_{\perp}\|^2 
- 2c \mathbf{A}\cdot\boldsymbol{\gamma}\right) = 0
\,,\\
(\partial _t -r \partial _s ) (\boldsymbol{\mu}\cdot \boldsymbol{\gamma}) + \frac{c}{2a} \partial _s \|\boldsymbol{\mu}_{\perp}\|^2 = 0
\,,\\
(\partial _t -r \partial _s )(\boldsymbol{\mu}\cdot \mathbf{A})
= \frac{c}{a}(\boldsymbol{\mu}_{\perp} \times\boldsymbol{\mu}) \cdot \mathbf{A} = 0
\,,\\
(\partial _t -r \partial _s ) (\boldsymbol{\gamma}\cdot \mathbf{A})  
= \frac{c}{a} (\boldsymbol{\mu}_\perp \times \boldsymbol{\gamma} ) \cdot \mathbf{A}
\ne0
\,,\\
(\partial _t -r \partial _s ) \|\boldsymbol{\gamma}\|^2 
= -\frac{2c}{a} \boldsymbol{\gamma}_{\perp} \cdot \partial_s 
\boldsymbol{\mu}_{\perp}
\ne0
\,.
\end{aligned}
\right.
\end{equation}
Consequently, we have three independent conservation laws 
\begin{align}
C_1 = \int 
\left(\frac{c}{a} \|\boldsymbol{\mu}_{\perp}\|^2 
- 2c \mathbf{A}\cdot\boldsymbol{\gamma}\right)
\,ds
\,,\quad
C_2 = \int \boldsymbol{\mu} \cdot\boldsymbol{\gamma}
\,ds
\,,\quad
C_3 = \int 
\boldsymbol{\mu} \cdot \mathbf{A}
\,ds
\,.
\label{3cons-laws}
\end{align}
Using the expression of the affine Lie--Poisson Hamiltonian vector 
field in \eqref{LP-Ham-struct-SO3}, it follows that $X_{C_2}(\boldsymbol{\mu}, \boldsymbol{\gamma}) = (\partial _s \boldsymbol{\mu}, 
\partial_s\boldsymbol{\gamma})$, so its flow is uniform translations 
in the space variable $s$. Similarly, since $X_{C_3}(\boldsymbol{\mu}, \boldsymbol{\gamma}) = (\boldsymbol{\mu} \times \mathbf{A}, 
\boldsymbol{\gamma}\times \mathbf{A})$, one concludes that its flow 
is rigid rotations about $\mathbf{A}$. A direct verification shows
that $C_1, C_2, C_3$ are in involution under the affine Lie--Poisson 
bracket defined by the operator in the right hand side of 
\eqref{LP-Ham-struct-SO3}. In addition, note that the Hamiltonian 
\eqref{SO3Ham} can be expressed as $h=C_1-2rC_2$. 
\medskip

\noindent \textsf{Linear instability for ZCR $SO(3)$- and 
$SL(2,\mathbb{R})$-strands.} By the third equation in 
\eqref{cons-eqns}, the parallel component ${\mu}_{\|}:=\mathbf{A}\cdot\boldsymbol{\mu}$ is conserved along traveling wave characteristics, 
$(\partial _t -r \partial _s ){\mu}_{\|}=0$, so its linear dispersion relation is $\omega=-rk$.
We write the equations \eqref{two_so3_zcr_equn} in the form
\[
\left(\partial _t -r \partial _s  - 
\frac{c}{a}\boldsymbol{\mu}_{\perp} \times \right)\boldsymbol{\mu} - 
c\boldsymbol{\gamma} \times \mathbf{A}=0 \quad\text{and}\quad 
\left(\partial_t -r \partial _s - 
\frac{c}{a} \boldsymbol{\mu}_{\perp} \times\right) \boldsymbol{\gamma}
+ \frac{c}{a}\partial_s \boldsymbol{\mu}_{\perp}  =0.
\]
Linearizing around the equilibrium solutions $\boldsymbol{\mu}_e = m\mathbf{A}$ and $\boldsymbol{\gamma}_e = n\mathbf{A}$ with constants $m$ and $n$ yields the dispersion relation 
\begin{align}
(\omega+rk)^2\Big( (\omega+rk)^4 - (m^2 - 2an)(\omega+rk)^2- a^2(k^2-n^2) \Big) = 0
\label{fulldispersion-rel}
\end{align}
from which we find the branches of the dispersion relations for the parallel and transverse components. The parallel components are stable, and the transverse components have a low wavenumber band of stable solutions for $(m^2 - 2an)^2\ge4a^2(n^2-k^2)\ge0$. It remains to determine how the solutions of the full nonlinear system will interact with the unstable transverse linear modes at higher wave numbers for this class of equilibrium solutions. 
However, we expect that nonlinearity cannot saturate the linear instability, because the conserved quadratic form provided by the Hamiltonian \eqref{SO3Ham} is sign-indefinite, which precludes Lyapunov stability. 

An analogous computation for the non-compact ZCR 
$SL(2, \mathbb{R})$-strand for the equilibria with $m=0=n$ removes the middle term in \eqref{fulldispersion-rel} and changes the sign of the remaining $a^2k^2$ term. In that case, only the parallel components of the linear modes can propagate stably and the linear instability is not controlled by nonlinear terms. Thus, one may conjecture that the equilibrium solutions of all $G$-Strand equations on semisimple Lie algebras that possess a quadratic ZCR will be linearly unstable, for Lie algebras with either compact or non-compact real forms. \quad 
$\blacklozenge$

\paragraph{Example: $ \mathfrak{g}  = \mathfrak{so}(4)$.} Implementing
the Lie algebra isomorphism
\[
\mathbb{R}^3 \times \mathbb{R}^3\ni (\mathbf{x}, \mathbf{y}) \longmapsto \frac{1}{2}\begin{bmatrix}
\;\widehat{\mathbf{x}}+ \widehat{\mathbf{y}} & 
\mathbf{x} - \mathbf{y}\\
-(\mathbf{x} - \mathbf{y})^\mathsf{T}& 0
\end{bmatrix} \in \mathfrak{so}(4)
\]
it immediately follows that the ZCR equations for $\mathfrak{so}(4)$
decouple into two ZCR equations for $SO(3)$. \quad 
$\blacklozenge$

\section{Conclusion and outlook}
Motivated partly by previous work on the zero curvature representation (ZCR) of completely integrable chiral models and partly by the underlying Hamiltonian structures of ideal complex fluids corresponding to \eqref{GS_Ham}, we have studied the class of $G$-Strand equations \eqref{GS_kappa_eqn} on semisimple Lie algebras that admit a ZCR that is quadratic in the spectral parameter. The ZCR conditions impose functional relations among the variables which reduce the dimension of the dynamics, but still preserve the Hamiltonian structure of the original equations. The main results from this investigation are the following.
\begin{enumerate}
\item
Using the root space decomposition, the equations for the $G$-Strand were formulated explicitly for any semisimple Lie algebra.
\item
ZCRs were constructed that are quadratic in a spectral parameter for two classes of equations. These classes follow from either a normal real form discussed in section \ref{case1-sec}, or a compact real form discussed in section \ref{case2-sec}. 
\item
The equilibria for these equations were studied for $\mathfrak{so}(3)$ and $\mathfrak{sl}(2,\mathbb{R})$, and were found to be linearly unstable for both classes of equations at sufficiently high wave numbers. The effect of these instabilities on the full nonlinear solution behavior in physically reasonable situations remains to be understood, perhaps through numerical simulations.
\end{enumerate}
As mentioned earlier, this work was partly motivated by the universal form of the affine Lie--Poisson bracket for ideal complex fluids \cite{Ho2002,GBRa2009} which, for semisimple Lie groups, leads to a quadratic ZCR. However, the usual physical applications of the theory of complex fluids tend to be rather dissipative, and inertial (Hamiltonian) terms are usually neglected. We hope the present result that the equilibria for the underlying ideal theory are typically unstable for semisimple symmetry groups will stimulate additional  investigations of the interactions between inertial and dissipative terms in complex fluid equations. 
\paragraph{Acknowledgments.}
We are grateful for hospitality at the Isaac Newton Institute for Mathematical Sciences, where this work was performed during its Complex Fluids program in 2013.
We thank our friends L. O'N\'{a}reigh, A. Rey, and C. Tronci for their kind encouragement and thoughtful remarks during the course of this work.


\begin{thebibliography}{}


\bibitem{Cahn1984}
Cahn, R. N. [1984] \textit{Semi-Simple Lie Algebras and
Their Representations}, The Benjamin/Cummings Publishing 
Company, Advanced Book Program, Menlo Park, California, USA, 1984.

\bibitem{Di1983a}
Dickey, L. A. [1983] 
Symplectic structure, Lagrangian, and involutiveness of first integrals of the principal chiral field equation. \textit{Commun, Math. Phys.} 87, 505--513 

\bibitem{Di1983b}
Dickey, L. A. [1983] 
Hamiltonian structures and Lax equations generated by matrix differential operators 
with polynomial dependence on a parameter, \textit{Commun. Math. Phys.} 88, 27--42. 

\bibitem{DuFoNo1992} 
Dubrovin, B. A., Fomenko, A. T., Novikov, S. P. ,  [1992]
\textit{Modern Geometry - Methods and Applications: Part I: The Geometry of Surfaces, Transformation Groups, and Fields}, 
Translated by R. G. Burns, Vol 93, Springer Graduate Texts in Mathematics, 1992.

\bibitem{DuKrNo1985} 
Dubrovin, B. A., Krichever, I. M., Novikov, S. P. 1990
Integrable systems I.
In \textit{Dynamical Systems IV}, V. I. Arnold and S. P. Novikov, editors, Springer Encyclopedia Math. Sci., Berlin, 1990.

\bibitem{EGBHPR2010}
Ellis, D. C. P., Gay-Balmaz, F., Holm, D. D., Putkaradze, V. and Ratiu, T. S. [2010]
Dynamics of charged molecular strands.
{\it Arch Rat Mech Anal} {197},  811--902.
\\ (Preprint at  \url{www.arxiv.org/abs/0901.2959})

\bibitem{FaTa1987}
Faddeev, L. D., Takhtajan, L. A. [1987] \textit{Hamiltonian Methods in the Theory of Solitons}. Translated by A. G. Reyman, Springer-Verlag, Berlin. 

\bibitem{FoTr1988}  
Fomenko, A. T. and Trofimov, V. V. [1988] \textit{Integrable Systems on Lie Algebras and Symmetric Spaces}. Translated from the Russian by A. Karaulov, P. D. Rayfield, and A. Weisman, Advanced Studies in Contemporary Mathematics, \textbf{2}, Gordon and Breach Science Publishers, New York, 1988.

\bibitem{GBRa2009}
Gay-Balmaz, F. and Ratiu, T. S. [2009] The geometric structure of
complex fluids, {\it Adv. in Appl. Math.} 42: 2, 176--275.
\url{http://arxiv.org/pdf/0903.4294.pdf}


\bibitem{Ge2012} 
Gerdjikov, V. S. [2012] Riemann-Hilbert Problems with canonical normalization
 and families of commuting operators, {\it Pliska Stud. Math. Bulgar.} 21, 201-216.
 \\
(Preprint at   \url{http://arxiv.org/pdf/1204.2928v1.pdf})

\bibitem{Ge2013}
Gerdjikov, V. S.  [2013]
On new types of integrable 4-wave interactions
(Preprint at  \url{http://arxiv.org/pdf/1302.1116.pdf} 

\bibitem{Ho2002}
Holm, D. D. [2002] Euler-Poincar\'e dynamics of perfect complex fluids, in {\it Geometry, Dynamics and Mechanics: 60th Birthday Volume for J.E. Marsden}. P. Holmes, P. Newton, and A. Weinstein, eds., Springer-Verlag, pp 113--167.
\url{http://xxx.lanl.gov/abs/nlin.CD/0103041}

\bibitem{Ho2011}  Holm, D. D.  [2011] \emph{Geometric Mechanics, Part 2}, 2nd Edition, Chapter 10, Imperial College Press. 

\bibitem{HoIv2013}
Holm, D. D. and Ivanov, R. I. [2013] Matrix $G$-Strands, submitted June 2013 to \textit{Nonlinearity}.
(Preprint at  \url{arXiv:1305.4010})

\bibitem{HoIvPe2012}
Holm, D. D., Ivanov, R. I. and Percival, J. R. [2012] $G$-Strands, \textit{J. Nonlinear Sci.} 22: 517--551. (Preprint at  \url{arXiv:1109.4421})

\bibitem{HoMaRa1998} 
Holm, D. D.,  Marsden, J. E. and Ratiu, T. S. [1998]
The Euler--Poincar\'e equations and semidirect products
with applications to continuum theories,
{\em Adv. in Math.} \textbf{137}, 1--81;
Ibid [1998] 
Euler--Poincar\'e models of ideal fluids with nonlinear dispersion
{\em Phys. Rev. Lett.} \textbf{349}, 4173--4177.


\bibitem{Humphreys1978}
Humphreys, J. E. [1978] \textit{Introduction to Lie algebras and Representation Theory}. Second printing, revised. Graduate Texts in Mathematics, \textbf{9}, Springer-Verlag, New York-Berlin, 1978.

\bibitem{Ma1994}
Ma\~{n}as, M.  [1994] The principal chiral model as an 
integrable system. In \textit{Harmonic Maps and Integrable 
Systems}, Fordy, A. P. and Wood, J. C., eds. pp. 147--173, 
Aspects of Mathematics, Volume E23, Vieweg, Braunschweig/Wiesbaden.

\bibitem{NoMaPiZa1984}
Novikov, S. P., Manakov, S. V., Pitaevski, I. P. and Zakharov, V. E. [1984] \textit{Theory of Solitons: The Inverse Scattering Method}. Consultants Bureau, New York.


\bibitem{Sternberg1983}
Sternberg, S. [1983] \textit{Lectures on Differential Geometry}. Second edition. With an appendix by Sternberg and Victor W. Guillemin. Chelsea Publishing Co., New York, 1983.

\bibitem{Uh1989} Uhlenbeck, K. [1989] Harmonic maps into Lie groups (Classical solutions of the chiral model). {\it J. Diff. Geom.} {\bf 30}, 1--50.


\bibitem{ZaMi1978} Zakharov, V. E. and Mikhailov, A. V. [1978] Relativistically invariant
two-dimensional models of field theory which are integrable by
means of the inverse scattering problem method. {\it Zh. Exp. Teor. Fiz.}
{\bf 74}, 1953--1973 (English translation: {\it Sov. Phys. JETP} {\bf 47},
1017--1027).

\bibitem{ZaMi1980} Zakharov, V. E. and Mikhailov, A. V. [1980] 
On the integrability of classical spinor models in two-dimensional space-time, 
{\it Commun. Math. Phys.} {\bf 74}, 21--40.

\end{thebibliography}
\end{document}